# Quantum transport properties of the topological Dirac Semimetal α-Sn

Md Shahin Alam [1,*], Alexandr Kazakov [1], Mujeeb Ahmad [1], Rajibul Islam [2], Fei Xue [2], and Marcin Matusiak [1,3,†]

1. International Research Centre MagTop, Institute of Physics, Polish Academy of Sciences, Aleja Lotników 32/46, PL-02668 Warsaw, Poland
2. Department of Physics, University of Alabama at Birmingham, Birmingham, Alabama 35294, USA
3. Institute of Low Temperature and Structure Research, Polish Academy of Sciences, ulica Okólna 2, 50-422 Wrocław, Poland

We report measurements of the electrical resistivity ($\rho$) and thermoelectric power ($S$) in a thin film of strained single-crystalline $\alpha$-Sn grown by molecular beam epitaxy on an insulating substrate. The temperature ($T$) dependence of the resistivity of $\alpha$-Sn can be divided into two regions: below $T^* \approx 135$ K $\rho(T)$ shows a metallic-like behaviour, while above this temperature an increasing contribution from thermally excited holes to electrical transport is observed. However, it is still dominated by highly mobile electrons, resulting in a negative sign of the Seebeck coefficient above $T = 47$ K. In the presence of the magnetic field ($B$) applied along an electric field or thermal gradient, we note a negative magnetoresistance or a negative slope of $S(B)$, respectively. The theoretical prediction for the former (calculated using density functional theory) agrees well with the experiment. However, these characteristics quickly disappear when the magnetic field is deviated from an orientation parallel to the electrical field or the thermal gradient. We indicate that the behaviour of the electrical resistivity and thermoelectric power can be explained in terms of the chiral current arising from the topologically non-trivial electronic structure of $\alpha$-Sn. Its decay at high temperature is a consequence of the decreasing ratio between the intervalley Weyl relaxation time to the Drude scattering time.

[*] shahin@magtop.ifpan.edu.pl, [†] m.matusiak@intibs.pl

# 1. Introduction

Over the past two decades, topological quantum materials have gained significant attention due to their nontrivial momentum-space topology [1–4]. For example, the Dirac semimetals, which could be considered three-dimensional analogues to the graphene, host four-fold-degenerate Dirac points protected by topological constraints [5–7]. Dirac nodes split into two chirally distinct Weyl nodes when at least one of the symmetries protecting the Dirac cone gets broken [8–10]. Dirac semimetals exhibit unique and exciting features such as ultra-high mobility [11], large magnetoresistance [11,12], chiral magnetic effect [13,14], etc. These exotic phenomena have been experimentally reported in a number of topological Dirac semimetals, namely $Na_3Bi$ [15,16], $Cd_2As_3$ [11,17,18], $ZrTe_5$ [19,20], $Bi_{1-x}Sb_x$ [21], $YbMnBi_2$ [22], TlBiSSe [12].

Recently, grey tin, or α-Sn, a zero-gap semiconductor emerges as an exciting material due to its nontrivial band topology [23–25]. It is the elemental candidate showing many topological phases, that can be tailored by various conditions, such as changing the thickness, imposing the strain, and applying the electric and magnetic fields [26,27]. The application of the in-plane tensile strain transforms α-Sn into a robust three-dimensional topological insulator with a large topological gap [24] and a high Fermi velocity [28], whereas in-plane compressive strain makes it a topological Dirac semimetal protected by four-fold rotational symmetry [26,27]. The in-plane compressive strain, which modifies the electronic structure of α-Sn, can be obtained experimentally by epitaxially growing a thin film on a substrate with lattice constants that do not exactly match those of grey tin. To obtain the Dirac semimetal phase, the desired mismatch is achieved by selecting the appropriate substrates, such as InSb(111) [29], InSb(001) [30], CdTe(111) [31],

GaAs(001) [32]. The degeneracy of the Dirac cones is lifted in the presence of an external magnetic field, turning α-Sn into a Weyl semimetal. The presence of a pair of Weyl points in momentum space can lead to appearance of a peculiar phenomenon when the electric and magnetic fields are applied parallel to each other. It is known as a chiral anomaly and, can be observed as a positive magneto-conductivity in real crystalline materials suggested by Nielsen-Ninomiya in 1983 [33]. Its origin is the charge pumping between Weyl points (WPs) of opposite chirality that occurs in a Weyl semimetal (WSM) subjected to the electric and magnetic fields applied in parallel along the direction of the WPs separation [13]. Such negative magnetoresistance has been reported by various groups in topological Dirac and Weyl semimetals [14,15,20,34,35], but there exist other mechanisms that could underlie this effect without involving charge pumping between the Weyl nodes [36–38]. In this regard, measurements of the thermoelectric effects offer an opportunity to complement to electrical measurements and gain valuable insight into electronic transport properties [39–42].

In this work, we investigate the electrical and thermoelectrical properties of the α-Sn thin film, which is a topological Dirac semimetal when deposited on a CdTe/GaAs (001) substrate. To ensure the best-quality, the sample was grown using molecular beam epitaxy (MBE). We report on the quadratic variation in the field dependencies of the electrical conductivity and thermopower, which can be explained as a consequence of the chiral anomaly. We also calculated the ratio of intervalley scattering time to mean free time, which indicates the role of chiral Weyl fermions in the emergence of the negative longitudinal magnetoresistance and the negative slope of thermopower at the high magnetic field.

## 2. Materials and methods

Epitaxial film of 200 nm thickness α-Sn was grown by the molecular beam epitaxy (MBE) on (001) GaAs substrate with a 4 μm CdTe buffer layer. CdTe buffer provides the necessary ~0.1 % compressive in-plane strain to induce the Dirac semimetal phase in grey tin. Extensive structural characterization confirmed the high quality of the obtained film without any inclusions of the metallic β-Sn phase, as well as the value of the strains. More details about the growth procedure, as well as the structural characterization of the studied film can be found in Reference [32].

To perform the electric transport measurements, we cut the sample with a length ($a$-axis) of 2.2 mm and a width ($b$-axis) of 0.6 mm. This reasonable aspect ratio of the sample was chosen to minimize the geometric effects on our experimental data. We measured the resistivity using the standard four probe method. The electrical contacts were made with 25 μm thick gold wires and DuPont 4929 silver paint. The current contacts were made along the entire width of a sample to minimize the possibility of the current jetting effect occurrence [43]. The dc electrical current was applied using a Keithley 6221 current source and voltages along the sample were measured with a Keithley 2182A nanovoltmeter.

To measure the thermoelectric properties, the sample was mounted between the two phosphor bronze clamps with two Cernox thermometers attached. These were used to determine the thermal gradient along the sample generated by a Micro-Measurements strain gage heater (10 kΩ resistance), which was connected to Keithley 6221 current source. Thermoelectric voltage data were collected by an EM Electronics A20a DC sub-nanovolt amplifier working in conjunction with a Keithley 2182A nanovoltmeter. The temperature dependence of thermoelectric power was measured with the heater off and on method, while for the magnetic field (± 14.5 T) sweeps the heater was continuously turned on.

The electronic structure calculations were carried out with the projector augmented wave approach within the density functional theory framework, utilizing the VASP package [44]. A plane-wave energy cut-off of 650 eV was employed in the study. We have performed the calculation using a meta-GGA approach, which is based on the modified Becke-Johnson (MBJ) exchange potential together with local density approximation for the correlation potential scheme with the parameter CMBJ = 1.215 in order to get the experimental band ordering [45]. The calculations of electronic structures were performed with a 12 ×12 ×10 Monkhorst-Pack k-mesh [46], incorporating the spin-orbit coupling self-consistently. The VASPWANNIER90 interface was employed in our study and we utilized s and p orbitals of Sn atoms to generate ab initio tight-binding Hamiltonian without performing the procedure for maximizing localization [47,48]. The calculation of the surface state was performed using the semi-infinite Green's function approach incorporated in the Wanniertools [49,50].

## 3. Results and Discussions

The thermoelectric power ($S$) and electrical resistivity ($\rho$) were measured respectively with the thermal gradient ($\nabla T$) or electrical current ($j$) applied along $a$-axis of the α-Sn thin film. Figure 1 presents the temperature dependence of resistivity in zero magnetic field ($B$) measured in the temperature range 2 - 300 K. The entire $\rho(T)$ dependence can be divided into two temperature regions, namely, for $T \gtrsim 135$ K we observe a semiconducting behaviour that can be attributed to the increasing with temperature contribution to electrical transport from thermally excited holes [51]. Those holes originate from the thermally driven transitions between valence and conduction parts of the $\Gamma_{8vc}^+$ band as well as from the indirect transition $L_{6c}^+ - \Gamma_{8vc}^+$ found in the calculated electronic structure of α-Sn [52]. In the latter, the magnitude of the band gap in the epitaxially stretched α-Sn thin film is temperature dependent and becomes larger as the temperature decreases [53]. In consequence, thermal hole excitation is prevented for $T \lesssim 135$ K and $\rho(T)$ behaves in a metallic-like manner. This is due to the electrons in the vicinity of Γ- bands dominating the transport features of α-Sn at low temperatures [51]. Interestingly, even at a very low temperature, some contribution to the electronic transport from hole-like charge carriers is still present [32].

It appears that the temperature dependence of the thermopower is also affected by the presence of two types of charge carriers. In almost the entire temperature range the thermopower is negative reflecting a dominating role of highly mobile electrons. The absolute value of $S$ reaches its maximum of 22.5 µV/K at $T \approx 150$ K. The upturn in $S(T)$ above this temperature likely reflects the discussed above increasing contribution from the thermally excited holes. For a multiband conductor, the total thermoelectric power is a sum of individual band contributions weighted by the respective conductivities. Within the two-band model this can be expressed as $S = (S_e \sigma_e + S_h \sigma_h)/\sigma$, where $\sigma = \sigma_e + \sigma_h$, where subscripts $e$ and $h$ denote contributions from electrons and holes, respectively. Therefore, at high temperatures the increasing positive participation from thermally excited holes will result in a decreasing absolute value of negative $S$ dominated by electrons. Below $T \approx 50$ K the Seebeck coefficient gets smaller than 1 µV/K and attains a positive value in the low temperature limit (see inset of the Fig. 1.), which can be also seen in the $S(B)$ dependences presented in Fig. 2b. This may be due to phonon-drag, but also due to contribution from the low-mobility holes, which were recently reported to be present at low temperatures in α-Sn thin films [32]. For the former S(T) should exhibit a maximum at the temperature of about $\Theta_D/5$ [54], where $\Theta_D$ is the Debye temperature. In grey tin $\Theta_D \approx 260$ K [55], but the maximum we observe is at the temperature somewhat lower than the expected $T \approx 50$ K. However, this is only an estimated position, which may be also effectively shifted by contribution from diffusive thermopower. Nevertheless, this positive contribution appears to be weakly dependent on the magnetic field, so does not affect our analysis.

The magnetoresistance of α-Sn was measured at different temperatures in the parallel configuration of the electrical current and magnetic field (j || B), the results are presented in Fig. 2a. Below $T \approx 100$ K, a Landau quantization occurs, leading to the appearance of pronounced Shubnikov de Haas (SdH) effect in high magnetic field. The resulting strong oscillations in ρ(B) are shown in Fig. S1(a) in Supplemental Material (SM) after subtraction of slowly varying background (3$^{rd}$ order polynomial). The fast Fourier transform (FFT) spectrum (see Fig. S1(b)) reveals a single slow frequency $F \sim 12$ T, that can be assigned to a bulk Dirac point near the fermi level of α-Sn, in agreement with the previous study [1]. The SdH oscillations can be described by the Lifshitz - Kosevich theory [56]:

$$\Delta\rho \propto R_T R_D R_S \cos\left[2\pi\left(\frac{F}{B} - \frac{1}{2} + \beta \pm \delta\right)\right], \quad (1)$$

The three damping factors are: the thermal reduction factor $R_T = \frac{\chi}{\sinh(\chi)}$, the Dingle damping factor $R_D = exp\left(-\chi \frac{T_D}{T}\right)$ ($T_D$ is the Dingle temperature) and the spin splitting term $R_S = \cos\left(\frac{p\pi}{2}\frac{gm^*}{m_e}\right)$. The parameter $\chi = \frac{2\pi^2 k_B T m^*/m_e}{e\hbar B}$, where $k_B = 1.381 \times 10^{-23}$ JK$^{-1}$ is the Boltzman constant, $e = 1.602 \times 10^{-19}$ C, $m_e = 9.108 \times 10^{-31}$ kg are the electron charge and mass, $\hbar = 1.054 \times 10^{-34}$ Js is reduced Planck's constant, and $m^*$ is the cyclotron mass of the charge carrier. Within the $R_S$ term, g is the Landé factor, p is the harmonic order. Under cosine function in Eqn. 1, F is the frequency of quantum oscillations (QO), $2\pi\beta$ is the

Berry phase, $\delta = \pm \frac{1}{8}$ is the phase shift related to the dimensionality of the electronic structure of a 3D system.

The effective mass can be calculated from the temperature dependence of the amplitude of the oscillations described by the thermal reduction factor $R_T$ (see Fig. S1(c) in the SM) and the resulting small effective mass $m^* \approx 0.01\ m_e$ is close to that previously reported [32]. The frequency of the quantum oscillations is directly related to the Fermi surface cross section area via the Onsager relation: $F = \left(\frac{\hbar}{2\pi e}\right) A_F$, where $A_F$ is the Fermi surface cross section area. Since the Fermi momentum $k_F = \sqrt{\frac{A_F}{\pi}}$, we have deduced the Fermi velocity $v_F = \frac{\hbar k_F}{m^*}$ and the Fermi energy $E_F = m^* v_F^2$ under the assumption that charge carriers have linear energy dispersion [57]. The estimated $v_F$ and $E_F$ are about $2.2 \times 10^6$ m/s and 277 meV respectively. Fig. S1(d) in SM shows that the experimental data can be well modelled with Eqn. 1 from which another parameter that can be obtained from the analysis of QO is the Dingle temperature, which was estimated to be $T_D$ = 11 K at 9.9 K temperature, again in good agreement with results reported in Ref. [32]. From the Dingle temperature, we can calculate the quantum life time $\tau_q = \frac{\hbar}{2\pi k_B T_D}$ and quantum mobility $\mu_q = \frac{e\tau_q}{m^*}$ [58], which are $\tau_q \approx 1 \times 10^{-13}$ s and $\mu_q \approx 5900$ cm$^2$V$^{-1}$s$^{-1}$, respectively. Such a high value of $\mu_q$ confirms the excellent transport properties of α-Sn.

Next, we will focus on a non-oscillatory component in ρ(B) of α-Sn, which, for j || B, exhibits a large negative longitudinal magnetoresistance (NLMR) at the high field and low temperature. With increasing temperature, NLMR decreases and vanishes above T ≈ 200 K. Intriguingly, the presence of NLMR for *j* || *B* was also reported in early studies of quantum oscillations in the bulk single crystals of α-Sn [59]. Furthermore, the authors observed an

additional 1/8 phase shift of the SdH oscillation, which is the expected result for the extra Berry phase of carriers from 3D band with linear dispersion [60]. These experimental facts initially remained uninterpreted, but have recently been linked to the presence of Dirac cones [26]. This is interesting because α-Sn in its pristine form should be, as mentioned above, a semiconductor with conduction and valence bands touching each other at the vertices. Perhaps even a small perturbation, such a presence of an internal strain, or a strain applied unwillingly during the experiment, can slightly shift the electronic bands, creating a non-trivial electron structure in the bulk sample [61].

The appearance of NLMR was reported for a large number of Dirac and Weyl semimetals [15,20,34,35,62] and it has been ascribed to pumping of the chiral charge between the Weyl nodes. On the other hand, questions were also raised as to whether the effect could be due to extrinsic mechanisms [36,38]. In particular, the current jetting contribution in highly mobile Weyl semimetals caused concerns about the detection of the chiral current in electrical measurements [43]. Therefore, the complementary measurements of other phenomena, such as the thermoelectric power, can provide a unique opportunity to investigate the effects of the chiral anomaly without current jetting artefact [41,43].

Figure 2b presents the magnetic field dependences of the thermopower (with $B \parallel \nabla T$) for selected temperatures. As mentioned above, in addition to the negative thermopower attributable to electrons, we also observe a roughly field independent positive contribution to the Seebeck coefficient. This appears in $S(B)$ data in Fig. 2b as a temperature dependent vertical shift that does not affect the field dependent part. At a low magnetic field, $S$ initially increases with $B$ – such a dependence was also observed in other topological Dirac semimetals and it was attributed to the process of Weyl nodes creation [43,63]. Below

$T \approx 50$ K we see a small peak in $S(B)$ at $B \approx 5.2$ T, which can be attributed to the quantum oscillations. For the frequency obtained from the SdH effect ($F = 12$ T), the next peak in $S(B)$ is expected at $B \approx 8.8$ T, which coincides with the maximum that occurs at $B \approx 9$ T. The contribution from QO to the total value of $S$ at the maximum can be calculated using the Lifshitz – Kosevich formula and is presented in Fig. S2 in the SM. At $T \approx 30$ K, the peak value of the QO from the baseline is $\approx 0.026$ $\mu VK^{-2}$, which is $\approx 33$ % of the total weight of $S$ at $B \approx 9$ T. However, the calculated contribution from QO significantly decreases with increasing temperature. For example, this becomes approximately 8 % at $T \approx 50$ K and vanishes completely above $T \gtrsim 55$ K, while the peak of $S(B)$ linked to formations of Weyl points is still present in the data. The maximum in $S(B)$ shifts to higher magnetic fields and at $B \approx 12$ T is clearly visible even at $T \approx 150$ K (see Fig. 2b). Thus, it is unlikely that the negative slope of $S(B)$ observed at high magnetic field originates from quantum oscillations and, alternatively, this type of anomalous field dependence was indicated as a manifestation of the chiral anomaly in several Dirac and Weyl semimetals [43].

To support our experimental observation, we have performed first-principles calculations using VASP [44] and Wannier90 [47]. Fig. 3(a) shows electronic band structure in the presence of spin orbit coupling (SOC), where the most of electronic bands are contributed by s and p orbital of Sn atoms. We observe a four-fold degenerate band crossing along Γ → Z direction with a camel back features [64,65](see Fig. 3(b)). The Dirac points are located 9 meV above the Fermi level $E_{Dirac} = E_f - 0.009$ eV at $(0, 0, \pm k_z)$, where $k_z = 0.398$ Å$^{-1}$, although it should be mentioned that density functional theory calculations can over- or underestimate the Fermi energy.

The pair of Dirac point can be shown in the electronic band structure in the 2D plane shown in Fig. 3(d). The application of magnetic field gives rise to a negative magneto-resistance demonstrated in the calculations when θ = 0° as shown in Fig. 3(e), which is consistent with experimental observation. Furthermore, we demonstrate the Dirac semimetal phase by investigating the band structure projected in Fig. S3 of the SM, (a) along the (100) direction and (b) along the (001) direction. The presence of topological surface state is evident with a bulk band crossing points. In addition, we calculate the Fermi surface projected on (100) and (001) surface in the bottom panel of Fig. S3. The Fermi surface along (100) shows the presence of close topological Fermi arc $k_y$-$k_z$ plane. These are $C_{4z}$ rotational symmetry-protected Dirac nodes that are against gap formation. In the presence of a finite Zeeman field that breaks time-reversal symmetry, each Dirac node separates into two Weyl nodes with the opposite chirality. Notably, these paired Weyl nodes are still aligned with the high symmetry direction protected by the crystal symmetry $C_{4z}$. The development of the chiral anomaly is facilitated by the combination of this property and the chiral nature of their lowest Landau [27,43].

If the negative slopes of ρ(B) and S(B) in the high field are in fact signs of the chiral anomaly, they should disappear when the magnetic field is tilted away from being parallel to *j* or ∇*T*, respectively. The angular dependences of longitudinal transport coefficients at *T* = 60 K are presented in Fig. 4. In the measurement of the magnetoresistance (Fig. 4a) we observed the maximal NLMR when $θ_ρ$ = 0° ($θ_ρ$ is the angle between *B* and *j*), whereas rotation of the magnetic field from in-plane to out-of-plane (from *a*- toward *c*-axis) quickly causes the dip in ρ(B) to vanish. Similar behaviour has been previously observed in topological semimetals, owing to the presence of a chiral current in the system [15,62,66]. The positive magnetoresistance for $θ_ρ$ ≳ 2.5° is due to the vanishing of the chiral anomaly influence and restoration of the orbital effect induced by the Lorentz force. Remarkably, the chiral anomaly appears to affect the magneto-thermopower (Fig. 4b) in a similar manner. Namely, the negative slope of $S(B)$ that occurs at a high magnetic field when $\theta_S = 0°$ ($\theta_S$ is the angle between $B$ and $\nabla T$) becomes positive when $B$ is away from $\nabla T$. The threshold for $\theta_S$, which is about 10°, is larger than that of $\theta_\rho$, possibly because the relative contribution from out-of-plane positive magneto-thermopower is not as large as that from orbital magnetoresistance.

The magnetic field breaks the time reversal symmetry, leading to the formation of Weyl nodes in α-Sn through the degeneracy of the Dirac nodes. In the presence of parallel electric and magnetic fields, the imbalance in the number of Weyl fermions of different chirality leads to the generation of an additional current that contributes to the total electrical conductivity. In the semi-classical regime, this can be expressed as [67,68]:

$$\sigma(B) = \sigma(0) + \sigma(0) * \frac{1}{3}\frac{\tau_i}{\tau}\frac{B^2}{B_q^2}, \qquad (2)$$

where $\sigma(0)$ is the Drude conductivity, $B_q = \frac{2E_F^2}{3e\hbar v_F^2}$ is the quantum magnetic field, and $\frac{\tau_i}{\tau}$ is the ratio of intervalley scattering time to the mean free time of charge carriers. The corresponding influence of the chiral anomaly on the thermoelectric power can be calculated using the Mott relation, which should obeyed if the chemical potential is larger than $k_BT$ and the scattering is dominated by elastic processes [41]. If the mean free time is assumed to be independent of energy, the equation reads [67,68]:

$$S(B) = S(0) - S(0) * 2\frac{\tau_i B^2}{3\tau B_q^2}, \qquad (3)$$

where $S(0)$ is the background thermopower unrelated to the chiral anomaly. Since the magnetic field can change the distance in k-space between Weyl cones, e.g. [69], the intervalley scattering time can in principle be expected to be field dependent. On the other hand, the model we used to describe the experimental data assumes $\tau_i$ to be field independent, which appears to be a sufficient approximation in the high field limit.

Moreover, this is in agreement with optical studies of the Dirac semimetal $Cd_3As_2$, which concluded that chiral relaxation shows little field dependence [70]. Equations 2 and 3 describe the behaviour of the non-oscillatory part of the respective signal, but this has appeared to be difficult to extract from field dependences of conductivity at low temperatures. This was done by simulating the SdH oscillations using Lifshitz - Kosevich formula (see Fig. S4 in SM) up to *B* = 8 T and then extrapolating the oscillatory signal to higher fields. Finally, we subtract the oscillations in order to obtain the non-oscillatory part of $\sigma(B)$, However, the results of extrapolation were not perfect below *T* = 100 K, hence we have chosen to restrict the upper limit of the fit to 10 T. As shown in Fig. 5a, $\sigma(B)$ at this intermediate field follows a quadratic dependence on *B* expected for the chiral anomaly [41,67,68] and can be approximated with Eqn. 2. Remarkably, the thermopower at a high magnetic field also appears to be well described by the model involving the chiral anomaly. Equation 3 predicts that the negative slope of the magneto-thermopower to be $B^2$ dependent component, which in fact appears in *S*(*B*) dependences – see Fig. 5b. From the fitting of the magnetoresistance and magneto-thermopower with Eqns. 2 and 3, we can estimate the ratio of the intervalley lifetime to the transport lifetime $\frac{\tau_i}{\tau}$, which is the parameter that defines whether an electronic system is in fact in the chiral limit [71]. The essential condition that must be met to observe the pumping of Weyl fermions from one node to another is $\tau_i > \tau$. Figure 6 presents the temperature dependences of $\frac{\tau_i}{\tau}$ calculated from $\sigma(B)$ and *S*(*B*), the latter multiplied by a prefactor 3/2, the reason for that will be discussed later. The agreement between these two independent experimental results is good i.e. the ratio $\frac{\tau_i}{\tau}$ decays with temperature, but both $\frac{\tau_i}{\tau}(T)$ do not coincide exactly. Similar behaviour was also reported for the topological Dirac semimetal $ZrTe_5$ [67]. For α-Sn, the relaxation time ratio from thermoelectric data is calculated with Eqn. 3 is smaller than one calculated from the electrical measurements. A likely reason for this discrepancy may be the approximations made to derive Eqn. 3. The first assumption is the energy (*E*) independent of carrier lifetime [67,68], which is usually not the case for real materials. The term in the Mott relations $\frac{d \ln \tau(E)}{dE}$ has significant contributions to the diffusion thermoelectric power [54]. For example, for metals at high temperature, the relation is expected to be $\tau(E) \propto E^{\frac{3}{2}}$ [72]. We include 3/2 factor in the relaxation time ratio calculated from thermoelectric data but our experimental results suggest even stronger energy dependence. This in fact was suggested to be significantly enhanced in the Dirac material SnTe [73]. Another important assumption in Eqn.3 was the strict validity of the Mott relation [67]. Since, the thermopower is measured under the condition of an imbalanced number, but possibly also energy, of chiral Weyl fermions [74] some deviations from the Mott relation can be expected. An actual reason will be an interesting subject of further investigation.

## 4. Conclusions

We studied the transport properties of the Dirac semimetal α-Sn in the configuration where the magnetic field is non-orthogonal to the electric field or thermal gradient. At the high field, we observed the negative longitudinal magnetoresistance and magneto-thermopower, which also shows a negative slope. Both features, which vary with field like $B^2$, can be attributed to the chiral anomaly and disappear in high temperatures. Further, a hint that we see manifestations of the chiral anomaly in α-Sn is a strong angular variation of the thermopower and resistivity. The calculated ratio of the intervalley scattering time to the mean free time satisfies the conditions for the chiral limit. We conclude that both the electrical and thermoelectric data indicate the presence of a Weyl system, which forms the chiral anomaly in the magnetic field.

**Acknowledgments**

We would like to thank V. Volobuiev for growing the sample and to K. Dybko for the discussion. This work was supported by the Foundation for Polish Science through the IRA Programme co-financed by EU within SG (Grant No. MAB/2017/1). R.I and F.X acknowledges the support by the National Science Foundation under Grant No. OIA-2229498. R.I. and F.X. acknowledge the access to the computing facility Cheaha at University of Alabama at Birmingham

**Competing financial interests:**

The authors declare no competing financial interests.

**Data availability**

All of the relevant data that support the findings of this study are available from the corresponding author upon reasonable request.

**M. S. Alam ORCID iD** : 0000-0001-9320-9150

**M. Matusiak ORCID iD**: 0000-0003-4480-9373

**Figures**

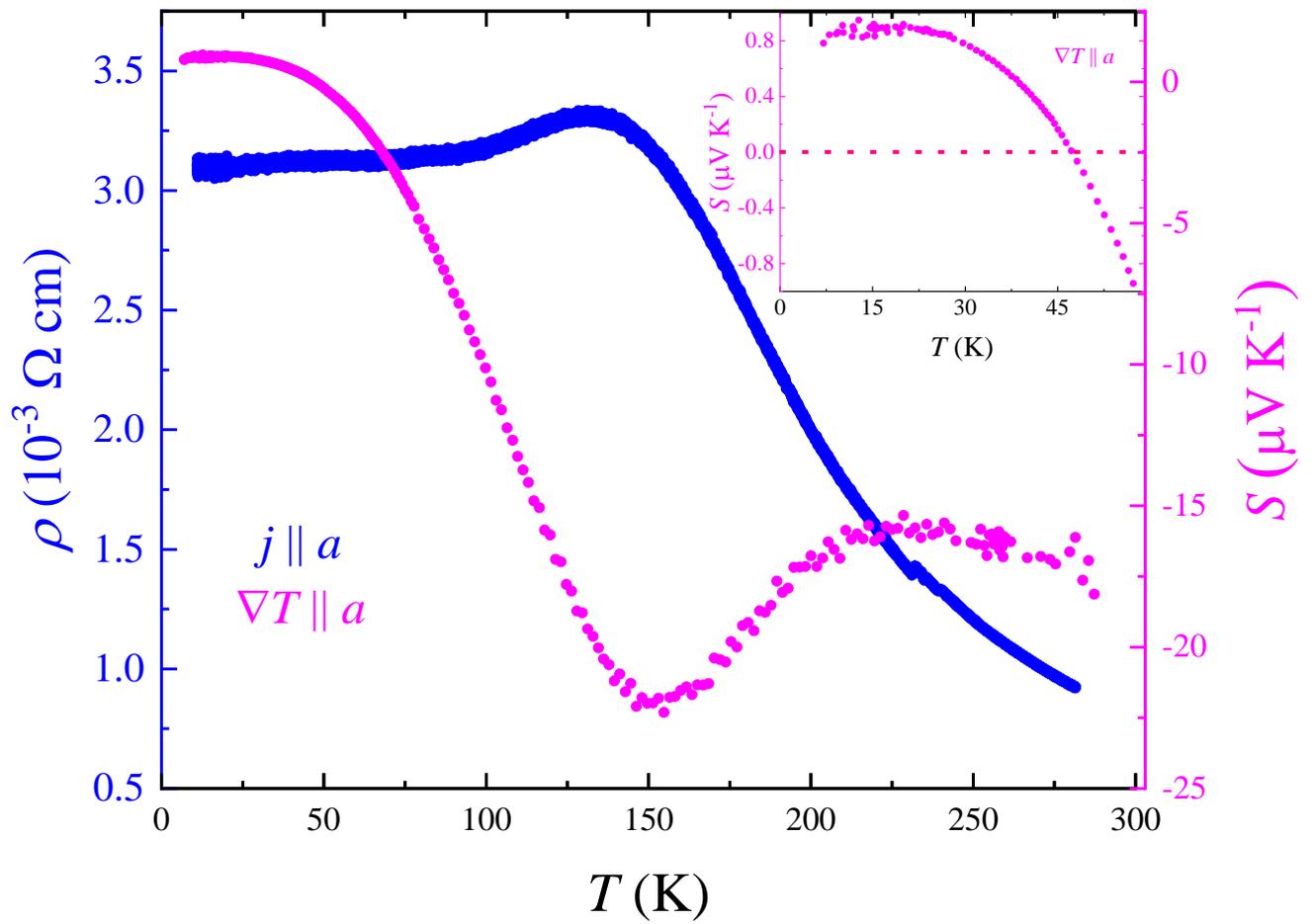

**Figure 1**. (Color online) Temperature dependences of the resistivity ($\rho$) and the thermoelectric power ($S$) of 200 nm thick $\alpha$-Sn thin film where the current ($J$) or thermal gradient ($\nabla T$) is applied parallel to $a$ – axis. Inset shows low temperature thermoelectric power data.

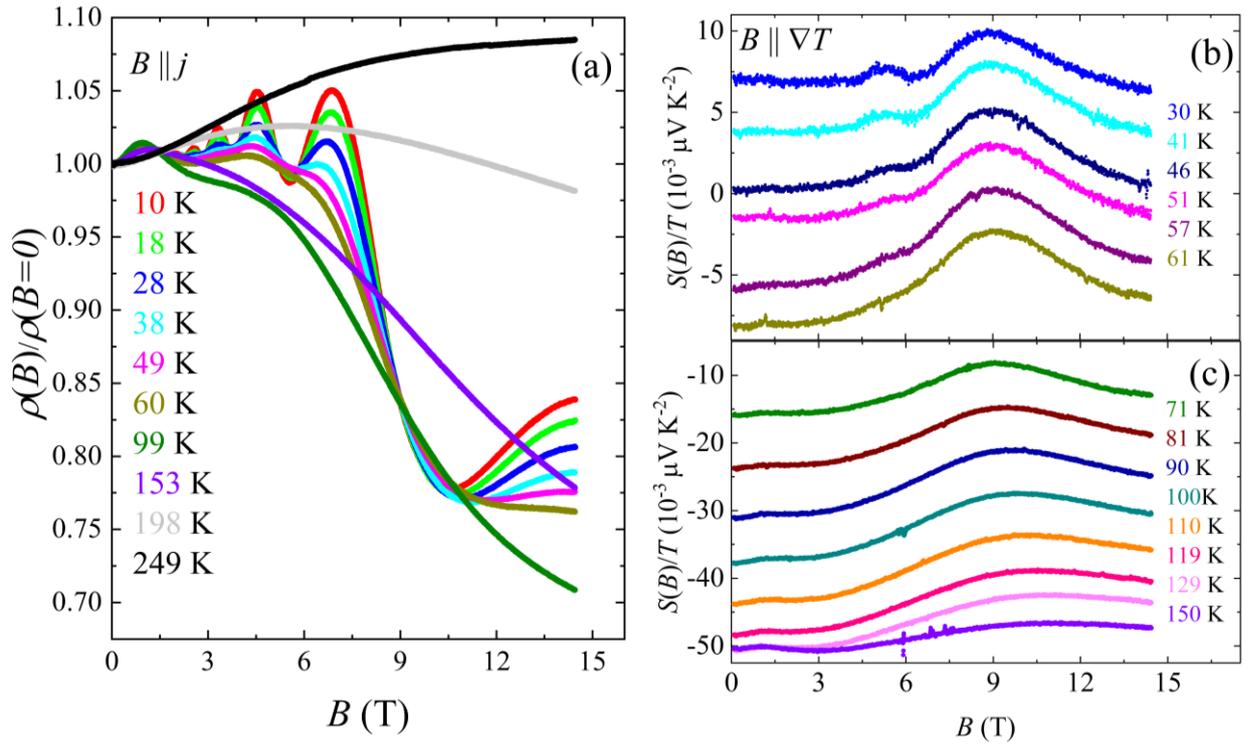

**Figure 2**. (Color online) Panel a: Normalized magnetoresistance versus magnetic field of α-Sn for selected temperatures when both magnetic field and current are applied parallel to *a*-axis (*B* || *j*). Magnetothermopower of α-Sn measured with the configuration of applied thermal gradient and magnetic field parallel to the *a*- axis (∇*T* || *B*), panel b: at low temperatures; panel c: at high temperatures.

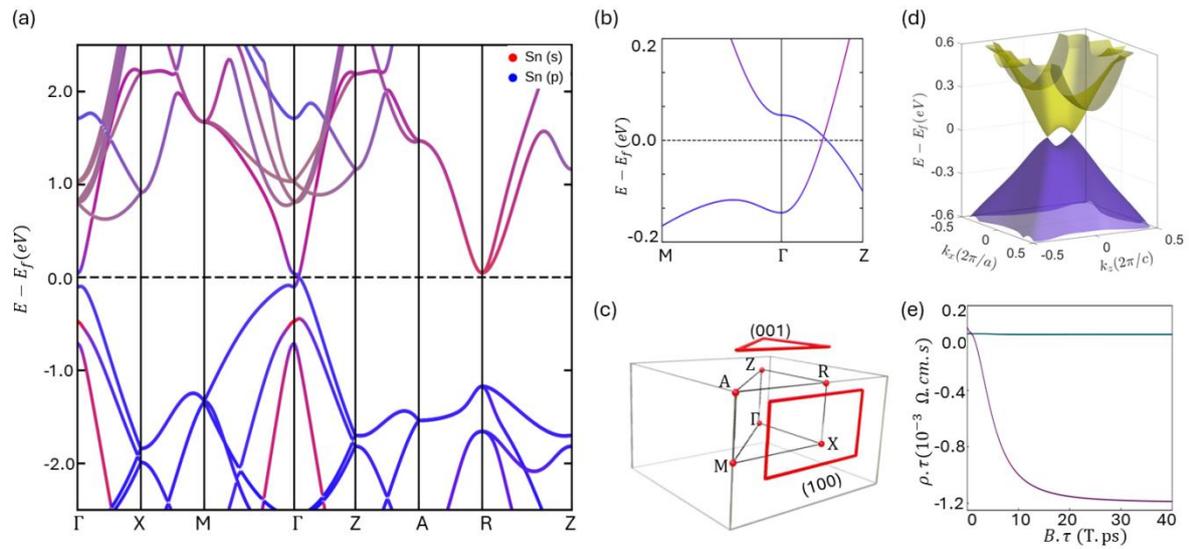

**Figure 3**. (Color online) a) The electronic band structure of α-Sn in the presence of spin-orbit coupling using density functional theory (DFT) over the full Brillouin zone (BZ); the BZ is shown in (c); the closer look of the along M→Γ→Z is shown in (b) . (d) The 2D band structure in the ($k_x$-$k_z$) plane illustrates the position of two Dirac points at (0,0,±$k_z$). (e) Magnetoresistance at θ = 0° magnetic field orientations, influenced by Fermi surface topology.

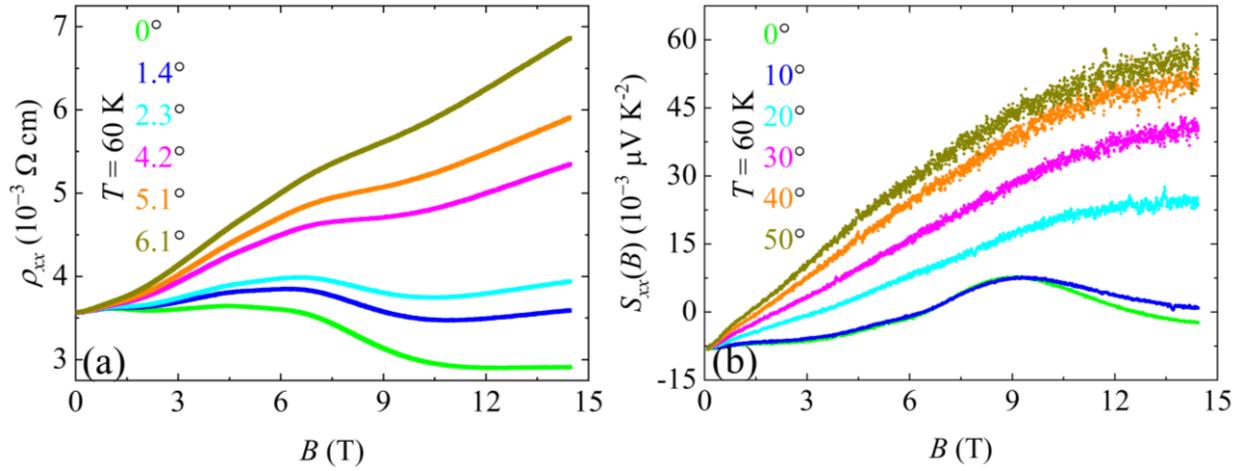

**Figure 4**. (Color online) Panel a: Resistivity ($\rho_{xx}$) in function of magnetic field ($B$) of Dirac semimetal α-Sn for selected angles (Θ, where Θ is the angle between $j$ and $B$) at a constant temperature 60 K. Panel b: Magnetothermopower ($S_{xx}(B)$) of α-Sn for selected angles (Θ, where Θ is the angle between $\nabla T$ and $B$) at a constant temperature 60 K.

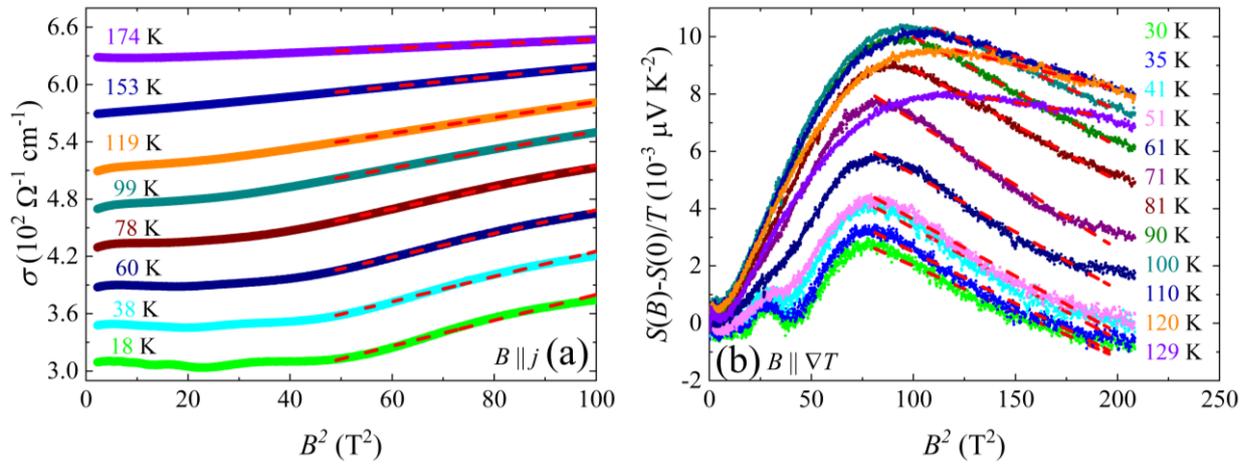

**Figure 5** . (Color online) Panel a: Conductivity (σ) in function of square of magnetic field (*B*) of α-Sn for several temperatures. For the sake of clarity, starting from $\sigma(B^2)$ for *T* = 38 K, the curves are successively shifted vertically by $10^2 \, \Omega^{-1} cm^{-1}$ each for sake of clarity. Panel b: Normalized thermopower *S(B)* of α-Sn for selected temperatures. The dashed line in both panels shows the fit as calculated from equations 2 and 3.

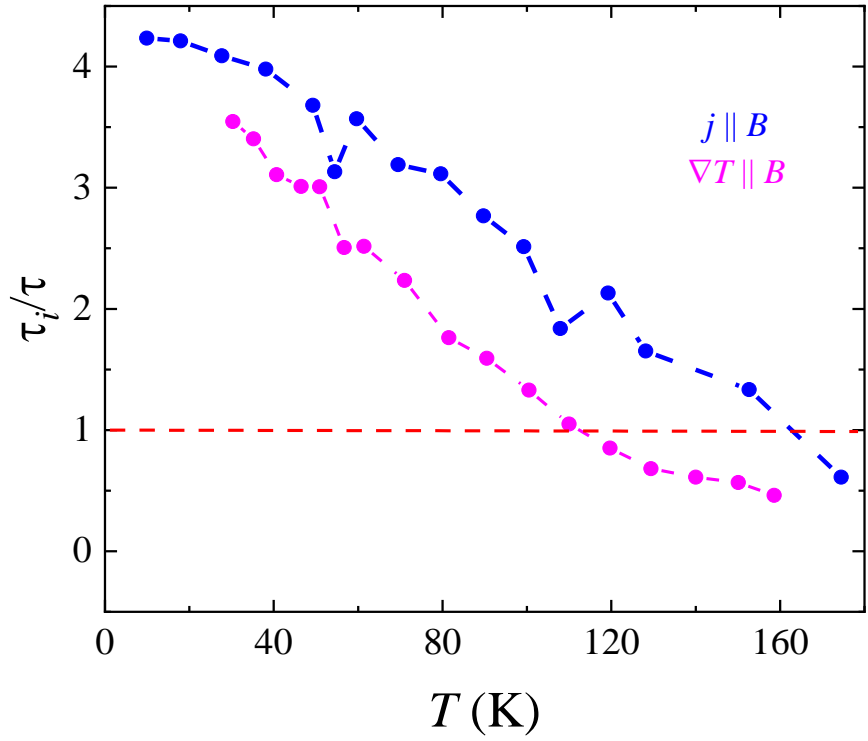

**Figure 6** . (Color online) The ratio of intervalley Weyl scattering time to Drude relaxation time ($\tau_i/\tau$) of $\alpha$-Sn sample in function of temperatures with the current ($j$) or thermal gradient ($\nabla T$) along with the magnetic field applied parallel to $a$ – axis.

# References

[1] C. L. Kane and E. J. Mele, *Quantum Spin Hall Effect in Graphene*, Phys. Rev. Lett. **95**, 226801 (2005).

[2] M. Z. Hasan and C. L. Kane, *Colloquium: Topological Insulators*, Rev. Mod. Phys. **82**, 3045 (2010).

[3] N. P. Armitage, E. J. Mele, and A. Vishwanath, *Weyl and Dirac Semimetals in Three-Dimensional Solids*, Rev. Mod. Phys. **90**, 015001 (2018).

[4] M. S. Alam, A. Fakhredine, M. Ahmad, P. K. Tanwar, H. Y. Yang, F. Tafti, G. Cuono, R. Islam, B. Singh, A. Lynnyk, C. Autieri, and M. Matusiak, *Sign Change of Anomalous Hall Effect and Anomalous Nernst Effect in the Weyl Semimetal CeAlSi*, Phys. Rev. B **107**, 85102 (2023).

[5] S. M. Young, S. Zaheer, J. C. Y. Teo, C. L. Kane, E. J. Mele, and A. M. Rappe, *Dirac Semimetal in Three Dimensions*, Phys. Rev. Lett. **108**, 140405 (2012).

[6] Z. K. Liu, B. Zhou, Y. Zhang, Z. J. Wang, H. M. Weng, D. Prabhakaran, S. Mo, Z. X. Shen, Z. Fang, X. Dai, Z. Hussain, and Y. L. Chen, *Topological Dirac Semimetal , $Na_3Bi$*, Science. **343**, 864 (2014).

[7] S. Borisenko, Q. Gibson, D. Evtushinsky, V. Zabolotnyy, B. Büchner, and R. J. Cava, *Experimental Realization of a Three-Dimensional Dirac Semimetal*, Phys. Rev. Lett. **113**, 027603 (2014).

[8] S. Y. Xu, I. Belopolski, N. Alidoust, M. Neupane, G. Bian, C. Zhang, R. Sankar, G. Chang, Z. Yuan, C. C. Lee, S. M. Huang, H. Zheng, J. Ma, D. S. Sanchez, B. K. Wang, A. Bansil, F. Chou, P. P. Shibayev, H. Lin, S. Jia, and M. Z. Hasan, *Discovery of a Weyl Fermion Semimetal and Topological Fermi Arcs*, Science. **349**, 613 (2015).

[9] S. Jia, S. Y. Xu, and M. Z. Hasan, *Weyl Semimetals, Fermi Arcs and Chiral Anomalies*, Nat. Mater. **15**, 1140 (2016).

[10] B. Yan and C. Felser, *Topological Materials: Weyl Semimetals*, Annu. Rev. Condens. Matter Phys. **8**, 337 (2017).

[11] T. Liang, Q. Gibson, M. N. Ali, M. Liu, R. J. Cava, and N. P. Ong, *Ultrahigh Mobility and Giant Magnetoresistance in the Dirac Semimetal $Cd_3As_2$*, Nat. Mater. **14**, 280 (2015).

[12] M. Novak, S. Sasaki, K. Segawa, and Y. Ando, *Large Linear Magnetoresistance in the Dirac Semimetal TlBiSSe*, Phys. Rev. B. Phys. **91**, 041203(R) (2015).

[13] A. A. Zyuzin and A. A. Burkov, *Topological Response in Weyl Semimetals and the Chiral Anomaly*, Phys. Rev. B. Phys. **86**, 115133 (2012).

[14] D. T. Son and B. Z. Spivak, *Chiral Anomaly and Classical Negative Magnetoresistance of Weyl Metals*, Phys. Rev. B. Phys. **88**, 104412 (2013).

[15] J. Xiong, S. K. Kushwaha, T. Liang, J. W. Krizan, M. Hirschberger, W. Wang, R. J. Cava, and N. P. Ong, *Evidence for the Chiral Anomaly in the Dirac Semimetal $Na_3Bi$*, Science. **350**, 413 (2015).

[16] S. Liang, J. Lin, S. Kushwaha, J. Xing, N. Ni, R. J. Cava, and N. P. Ong, *Experimental Tests of the Chiral Anomaly Magnetoresistance in the Dirac-Weyl Semimetals $Na_3Bi$ and GdPtBi*, Phys. Rev. X **8**, 031002 (2018).

[17] L. P. He, X. C. Hong, J. K. Dong, J. Pan, Z. Zhang, J. Zhang, and S. Y. Li, *Quantum Transport Evidence for the Three-Dimensional Dirac Semimetal Phase in $Cd_3As_2$*, Phys. Rev. Lett. **113**, 246402 (2014).

[18] M. Neupane, S. Y. Xu, R. Sankar, N. Alidoust, G. Bian, C. Liu, I. Belopolski, T. R. Chang, H. T. Jeng, H. Lin, A. Bansil, F. Chou, and M. Z. Hasan, *Observation of a Three-Dimensional Topological Dirac Semimetal Phase in High-Mobility $Cd_3As_2$*, Nat. Commun. **5**, 3786 (2014).

[19] G. Zheng, J. Lu, X. Zhu, W. Ning, Y. Han, H. Zhang, J. Zhang, C. Xi, J. Yang, H. Du, K. Yang, Y. Zhang, and M. Tian, *Transport Evidence for the Three-Dimensional Dirac Semimetal Phase in $ZrTe_5$*, Phys. Rev. B **93**, 115414 (2016).

[20] Q. Li, D. E. Kharzeev, C. Zhang, Y. Huang, I. Pletikosić, A. V. Fedorov, R. D. Zhong, J. A. Schneeloch, G. D. Gu, and T. Valla, *Chiral Magnetic Effect in $ZrTe_5$*, Nat. Phys. **12**, 550 (2016).

[21] H. J. Kim, K. S. Kim, J. F. Wang, M. Sasaki, N. Satoh, A. Ohnishi, M. Kitaura, M. Yang, and L. Li, *Dirac versus Weyl Fermions in Topological Insulators: Adler-Bell-Jackiw Anomaly in Transport Phenomena*, Phys. Rev. Lett. **111**, 246603 (2013).

[22] J. Y. Liu, J. Hu, Q. Zhang, D. Graf, H. B. Cao, S. M. A. Radmanesh, D. J. Adams, Y. L. Zhu, G. F. Cheng, X. Liu, W. A. Phelan, J. Wei, M. Jaime, F. Balakirev, D. A. Tennant, J. F. DItusa, I. Chiorescu, L. Spinu, and Z. Q. Mao, *A Magnetic Topological Semimetal $Sr_{1-y}Mn_{1-z}Sb_2$ (y, z < 0.1)*, Nat. Mater. **16**, 905 (2017).

[23] J. C. Rojas-Sánchez, S. Oyarzún, Y. Fu, A. Marty, C. Vergnaud, S. Gambarelli, L. Vila, M. Jamet, Y. Ohtsubo, A. Taleb-Ibrahimi, P. Le Fèvre, F. Bertran, N. Reyren, J. M. George, and A. Fert, *Spin to Charge Conversion at Room Temperature by Spin Pumping into a New Type of Topological Insulator: α -Sn Films*, Phys. Rev. Lett. **116**, 096602 (2016).

[24] Y. Ohtsubo, P. Le Fèvre, F. Bertran, and A. Taleb-Ibrahimi, *Dirac Cone with Helical Spin Polarization in Ultrathin α-Sn(001) Films*, Phys. Rev. Lett. **111**, 216401 (2013).

[25] M. T. Le Duc Anh, Kengo Takase, Takahiro Chiba, Yohei Kota, Kosuke Takiguchi, *Elemental Topological Dirac Semimetal α-Sn with High Quantum Mobility*, Adv. Mater **33**, 2104645 (2021).

[26] H. Huang and F. Liu, *Tensile Strained Gray Tin: Dirac Semimetal for Observing Negative Magnetoresistance with Shubnikov-de Haas Oscillations*, Phys. Rev. B **95**, 201101(R) (2017).

[27] D. Zhang, H. Wang, J. Ruan, G. Yao, and H. Zhang, *Engineering Topological Phases in the Luttinger Semimetal α -Sn*, Phys. Rev. B **97**, 195139 (2018).

[28] R. A. Carrasco, C. M. Zamarripa, S. Zollner, J. Menéndez, S. A. Chastang, J. Duan, G. J. Grzybowski, B. B. Claflin, and A. M. Kiefer, *The Direct Bandgap of Gray α -Tin Investigated by Infrared Ellipsometry*, Appl. Phys. Lett. **113**, 232104 (2018).

[29] V. A. Rogalev, F. Reis, F. Adler, M. Bauernfeind, J. Erhardt, A. Kowalewski, M. R. Scholz, L. Dudy, L. B. Duffy, T. Hesjedal, M. Hoesch, G. Bihlmayer, J. Schäfer, and R. Claessen, *Tailoring the Topological Surface State in Ultrathin α -Sn(111) Films*, Phys. Rev. B **100**, 245144 (2019).

[30] A. Barfuss, L. Dudy, M. R. Scholz, H. Roth, P. Höpfner, C. Blumenstein, G. Landolt, J. H. Dil, N. C. Plumb, M. Radovic, A. Bostwick, E. Rotenberg, A. Fleszar, G. Bihlmayer, D. Wortmann, G. Li, W. Hanke, R. Claessen, and J. Schäfer, *Elemental Topological Insulator with Tunable Fermi Level: Strained α-Sn on Insb(001)*, Phys. Rev. Lett. **111**, 157205 (2013).

[31]  S. Küfner, L. Matthes, and F. Bechstedt, *Quantum Spin Hall Effect in α-Sn/CdTe(001) Quantum-Well Structures*, Phys. Rev. B **93**, 045304 (2016).

[32]  J. Polaczyński, G. Krizman, A. Kazakov, B. Turowski, J. B. Ortiz, R. Rudniewski, T. Wojciechowski, P. Dłużewski, M. Aleszkiewicz, W. Zaleszczyk, B. Kurowska, Z. Muhammad, M. Rosmus, N. Olszowska, L.-A. De Vaulchier, Y. Guldner, T. Wojtowicz, and V. V. Volobuev, *3D Topological Semimetal Phases of Strained α-Sn on Insulating Substrate*, ArXiv:2309.03951 (2023).

[33]  H. B. Nielsen and M. Ninomiya, *The Adler-Bell-Jackiw Anomaly and Weyl Fermions in a Crystal*, Phys. Lett. B **130**, 389 (1983).

[34]  C. Z. Li, L. X. Wang, H. Liu, J. Wang, Z. M. Liao, and D. P. Yu, *Giant Negative Magnetoresistance Induced by the Chiral Anomaly in Individual $Cd_3As_2$ Nanowires*, Nat. Commun. **6**, 1 (2015).

[35]  P. K. Tanwar, M. Ahmad, M. S. Alam, X. Yao, F. Tafti, and M. Matusiak, *Gravitational Anomaly in the Ferrimagnetic Topological Weyl Semimetal NdAlSi*, Phys. Rev. B **108**, L161106 (2023).

[36]  K. Yoshida, *A Geometrical Transport Model for Inhomogeneous Current Distribution in Semimetals under High Magnetic Fields*, J. Phys. Soc. Jpn **40**, 1027 (1976).

[37]  K. Yoshida, *Anomalous Electric Fields in Semimetals under High Magnetic Fields*, J. Phys. Soc. Jpn **39**, 1443 (1975).

[38]  F. Arnold, C. Shekhar, S. C. Wu, Y. Sun, R. D. Dos Reis, N. Kumar, M. Naumann, M. O. Ajeesh, M. Schmidt, A. G. Grushin, J. H. Bardarson, M. Baenitz, D. Sokolov, H. Borrmann, M. Nicklas, C. Felser, E. Hassinger, and B. Yan, *Negative Magnetoresistance without Well-Defined Chirality in the Weyl Semimetal TaP*, Nat. Commun. **7**, 11615 (2016).

[39]  P. Wei, W. Bao, Y. Pu, C. N. Lau, and J. Shi, *Anomalous Thermoelectric Transport of Dirac Particles in Graphene*, Phys. Rev. Lett. **102**, 166808 (2009).

[40]  Y. M. Zuev, W. Chang, and P. Kim, *Thermoelectric and Magnetothermoelectric Transport Measurements of Graphene*, Phys. Rev. Lett. **102**, 096807 (2009).

[41]  B. Z. Spivak and A. V. Andreev, *Magnetotransport Phenomena Related to the Chiral Anomaly in Weyl Semimetals*, Phys. Rev. B **93**, 085107 (2016).

[42]  M. S. Alam, P. K. Tanwar, K. Dybko, A. S. Wadge, P. Iwanowski, A. Wiśniewski, and M. Matusiak, *Temperature-Driven Spin-Zero Effect in $TaAs_2$*, J. Phys. Chem. Solids **170**, 110939 (2022).

[43]  N. P. Ong and S. Liang, *Experimental Signatures of the Chiral Anomaly in Dirac–Weyl Semimetals*, Nat. Rev. Phys. **3**, 394 (2021).

[44]  G. Kresse and J. Furthmüller, *Efficient Iterative Schemes for Ab Initio Total-Energy Calculations Using a Plane-Wave Basis Set*, Phys. Rev. B **54**, 11169 (1996).

[45]  F. Tran and P. Blaha, *Accurate Band Gaps of Semiconductors and Insulators with a Semilocal Exchange-Correlation Potential*, Phys. Rev. Lett. **102**, 226401 (2009).

[46]  J. D. Pack and H. J. Monkhorst, *"special Points for Brillouin-Zone Integrations"*, Phys. Rev. B **13**, 5188 (1976).

[47]  A. A. Mostofi, J. R. Yates, Y. S. Lee, I. Souza, D. Vanderbilt, and N. Marzari, *Wannier90: A Tool for Obtaining Maximally-Localised Wannier Functions*, Comput. Phys. Commun. **178**, 685

(2008).

[48] N. Marzari and D. Vanderbilt, *Maximally Localized Generalized Wannier Functions for Composite Energy Bands*, Phys. Rev. B Phys. **56**, 12847 (1997).

[49] Q. S. Wu, S. N. Zhang, H. F. Song, M. Troyer, and A. A. Soluyanov, *WannierTools: An Open-Source Software Package for Novel Topological Materials*, Comput. Phys. Commun. **224**, 405 (2018).

[50] M. P. Lopez Sancho, J. M. Lopez Sancho, and J. Rubio, *Highly Convergent Schemes for the Calculation of Bulk and Surface Green Functions*, J. Phys. F Met. Phys. **15**, 851 (1985).

[51] G. de C. Owen Vail, Patrick Taylor, Patrick Folkes, Barbara Nichols, Brian Haidet, Kunal Mukherjee, *Growth and Magnetotransport in Thin-Film α-Sn on CdTe*, Phys. Status Solidi B **257**, 1800513 (2020).

[52] T. Brudevoll, D. S. Citrin, M. Cardona, and N. E. Christensen, *Electronic Structure of -Sn and Its Dependence on Hydrostatic Strain*, Phys. Rev. B **48**, 8629 (1993).

[53] C. A. Hoffman, J. R. Meyer, R. J. Wagner, F. J. Bartoli, M. A. Engelhardt, and H. Höchst, *Three-Band Transport and Cyclotron Resonance in -Sn and -Sn$_{1-x}$Ge$_x$ Grown by Molecular-Beam Epitaxy*, Phys. Rev. B **40**, 11693 (1989).

[54] D.K.C. MacDonald, Thermoelectricity: An Introduction to the Principles (Wiley, New York, 1962).

[55] N. Dementev, *Thermodynamical Insight on the Role of Additives in Shifting the Equilibrium between White and Grey Tin*, ArXiv:1011.2275 (2010).

[56] D. Shoenberg, *Magnetic Oscillations in Metals*, Magn. Oscil. Met. Cambridge Univ. Press (1984).

[57] Z. Ren, A. A. Taskin, S. Sasaki, K. Segawa, and Y. Ando, *Large Bulk Resistivity and Surface Quantum Oscillations in the Topological Insulator Bi$_2$T$_2$Se*, Phys. Rev. B **82**, 241306 (2010).

[58] A. Narayanan, M. D. Watson, S. F. Blake, N. Bruyant, L. Drigo, Y. L. Chen, D. Prabhakaran, B. Yan, C. Felser, T. Kong, P. C. Canfield, and A. I. Coldea, *Linear Magnetoresistance Caused by Mobility Fluctuations in n -Doped Cd$_3$As$_2$*, Phys. Rev. Lett. **114**, 117201 (2015).

[59] E. D. Hinkley and A. W. Ewald, *Oscillatory Magnetoresistance in Gray Tin*, Phys. Rev. **134**, A1261 (1964).

[60] C. M. Wang, H. Z. Lu, and S. Q. Shen, *Anomalous Phase Shift of Quantum Oscillations in 3D Topological Semimetals*, Phys. Rev. Lett. **117**, 077201 (2016).

[61] C. Z. Xu, Y. H. Chan, Y. Chen, P. Chen, X. Wang, C. Dejoie, M. H. Wong, J. A. Hlevyack, H. Ryu, H. Y. Kee, N. Tamura, M. Y. Chou, Z. Hussain, S. K. Mo, and T. C. Chiang, *Elemental Topological Dirac Semimetal: α -Sn on InSb(111)*, Phys. Rev. Lett. **118**, 146402 (2017).

[62] X. Huang, L. Zhao, Y. Long, P. Wang, D. Chen, Z. Yang, H. Liang, M. Xue, H. Weng, Z. Fang, X. Dai, and G. Chen, *Observation of the Chiral-Anomaly-Induced Negative Magnetoresistance: In 3D Weyl Semimetal TaAs*, Phys. Rev. X **5**, 031023 (2015).

[63] M. Hirschberger, S. Kushwaha, Z. Wang, Q. Gibson, S. Liang, C. A. Belvin, B. A. Bernevig, R. J. Cava, and N. P. Ong, *The Chiral Anomaly and Thermopower of Weyl Fermions in the Half-Heusler GdPtBi*, Nat. Mater. **15**, 1161 (2016).

[64] R. Islam, B. Ghosh, G. Cuono, A. Lau, W. Brzezicki, A. Bansil, A. Agarwal, B. Singh, T. Dietl, and C. Autieri, *Topological States in Superlattices of HgTe Class of Materials for Engineering Three-Dimensional Flat Bands*, Phys. Rev. Res. **4**, 023114 (2022).

[65] Z. Wang, H. Weng, Q. Wu, X. Dai, and Z. Fang, *Three-Dimensional Dirac Semimetal and Quantum Transport in $Cd_3As_2$*, Phys. Rev. B **88**, 125427 (2013).

[66] C. Zhang, E. Zhang, W. Wang, Y. Liu, Z. G. Chen, S. Lu, S. Liang, J. Cao, X. Yuan, L. Tang, Q. Li, C. Zhou, T. Gu, Y. Wu, J. Zou, and F. Xiu, *Room-Temperature Chiral Charge Pumping in Dirac Semimetals*, Nat. Commun. **8**, 13741 (2017).

[67] W. Zhang, P. Wang, G. Gu, X. Wu, and L. Zhang, *Negative Longitudinal Magnetothermopower in the Topological Semimetal $ZrTe_5$*, Phys. Rev. B **102**, 115147 (2020).

[68] Z. Jia, C. Li, X. Li, J. Shi, Z. Liao, D. Yu, and X. Wu, *Thermoelectric Signature of the Chiral Anomaly in $Cd_3As_2$*, Nat. Commun. **7**, 13013 (2016).

[69] F. Caglieris, C. Wuttke, S. Sykora, V. Süss, C. Shekhar, C. Felser, B. Büchner, and C. Hess, *Anomalous Nernst Effect and Field-Induced Lifshitz Transition in the Weyl Semimetals TaP and TaAs*, Phys. Rev. B **98**, 201107(R) (2018).

[70] B. Cheng, T. Schumann, S. Stemmer, and N. P. Armitage, *Probing Charge Pumping and Relaxation of the Chiral Anomaly in a Dirac Semimetal*, Sci. Adv. **7**, eabg0914 (2021).

[71] C. L. Zhang, S. Y. Xu, I. Belopolski, Z. Yuan, Z. Lin, B. Tong, G. Bian, N. Alidoust, C. C. Lee, S. M. Huang, T. R. Chang, G. Chang, C. H. Hsu, H. T. Jeng, M. Neupane, D. S. Sanchez, H. Zheng, J. Wang, H. Lin, C. Zhang, H. Z. Lu, S. Q. Shen, T. Neupert, M. Z. Hasan, and S. Jia, *Signatures of the Adler-Bell-Jackiw Chiral Anomaly in a Weyl Fermion Semimetal*, Nat. Commun. **7**, 10735 (2016).

[72] A. H. Wilson, *The theory of metals*, I. Proc. R. Soc. Lond. A **138**, 594 (1932).

[73] T. H. Liu, J. Zhou, M. Li, Z. Ding, Q. Song, B. Liao, L. Fu, and G. Chen, *Electron Mean-Free-Path Filtering in Dirac Material for Improved Thermoelectric Performance*, Proc. Natl. Acad. Sci. U. S. A. **115**, 879 (2018).

[74] J. Gooth, A. C. Niemann, T. Meng, A. G. Grushin, K. Landsteiner, B. Gotsmann, F. Menges, M. Schmidt, C. Shekhar, V. Süß, R. Hühne, B. Rellinghaus, C. Felser, B. Yan, and K. Nielsch, *Experimental Signatures of the Mixed Axial-Gravitational Anomaly in the Weyl Semimetal NbP*, Nature **547**, 324 (2017).

# Supplemental material:

## Quantum transport properties of the topological Dirac Semimetal α-Sn

Md Shahin Alam [1,*] Alexandr Kazakov [1], Mujeeb Ahmad [1], Rajibul Islam [2], Fei Xue [2], and Marcin Matusiak [1,3,†]

1. International Research Centre MagTop, Institute of Physics, Polish Academy of Sciences, Aleja Lotników 32/46, PL-02668 Warsaw, Poland
2. Department of Physics, University of Alabama at Birmingham, Birmingham, Alabama 35294, USA
3. Institute of Low Temperature and Structure Research, Polish Academy of Sciences, ulica Okólna 2, 50-422 Wrocław, Poland

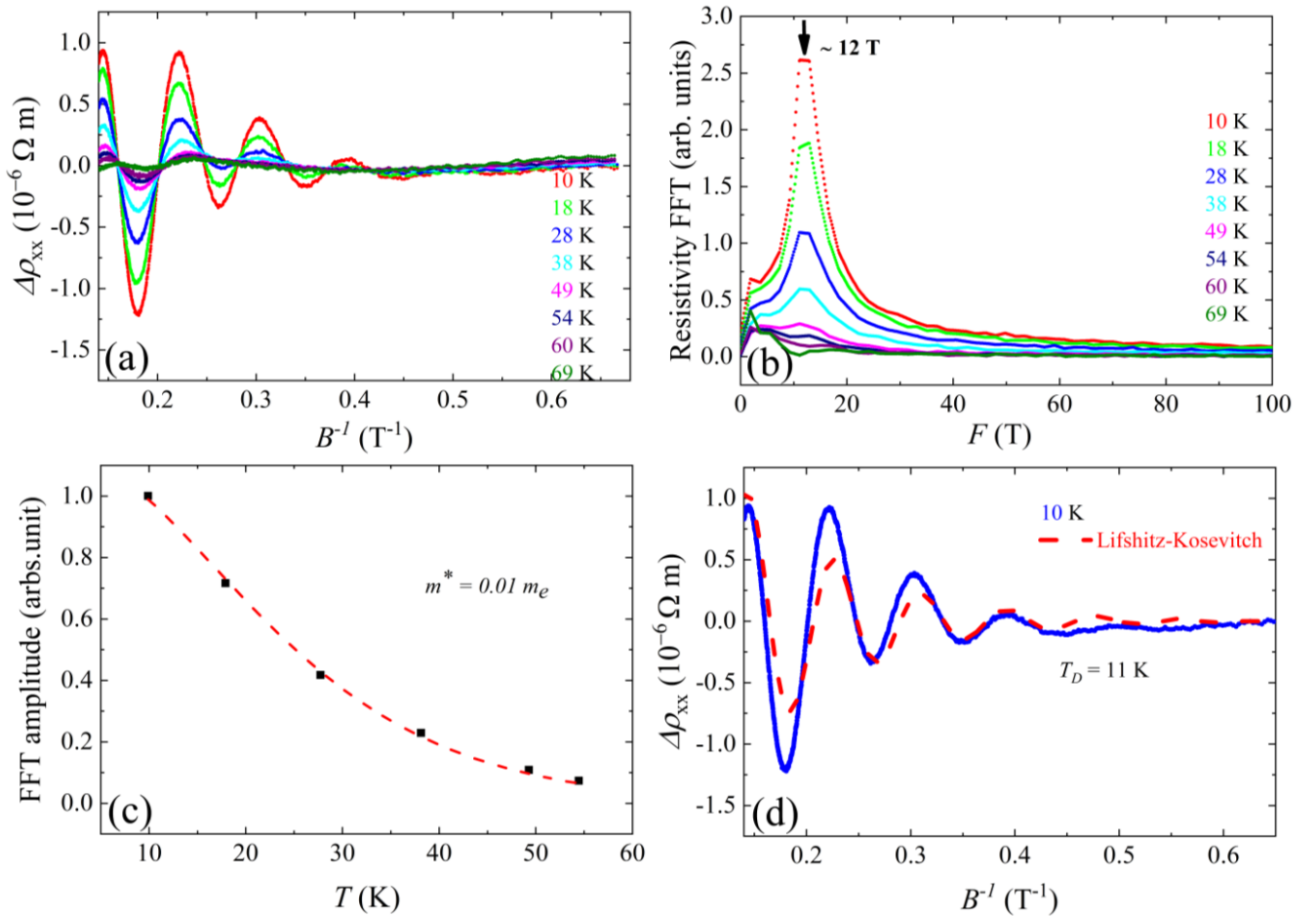

**Figure S1**. (Color online) (a) Shubnikov de Hass oscillations after subtracting the smooth 3rd order polynomial background from $\rho_{xx}(B)$ at different temperatures. (b) Fast Fourier transform spectrum of α-Sn showing a single frequency at ~12 T, calculated from SdH oscillations data, the effective field of oscillations taken between 1.5-8 T. (c) Amplitude of the FFT spectrum of the ~ 12 T frequency plotted against temperatures. Red dashes lines

represent the fit to the thermal damping factor ($R_T$) from the Lifshitz-Kosevich formula. (d) Shubnikov de Hass oscillations at 10 $K$ temperature. Red dashes line presents the fit to the Lifshitz-Kosevich formula (see equation 1 in the main text).

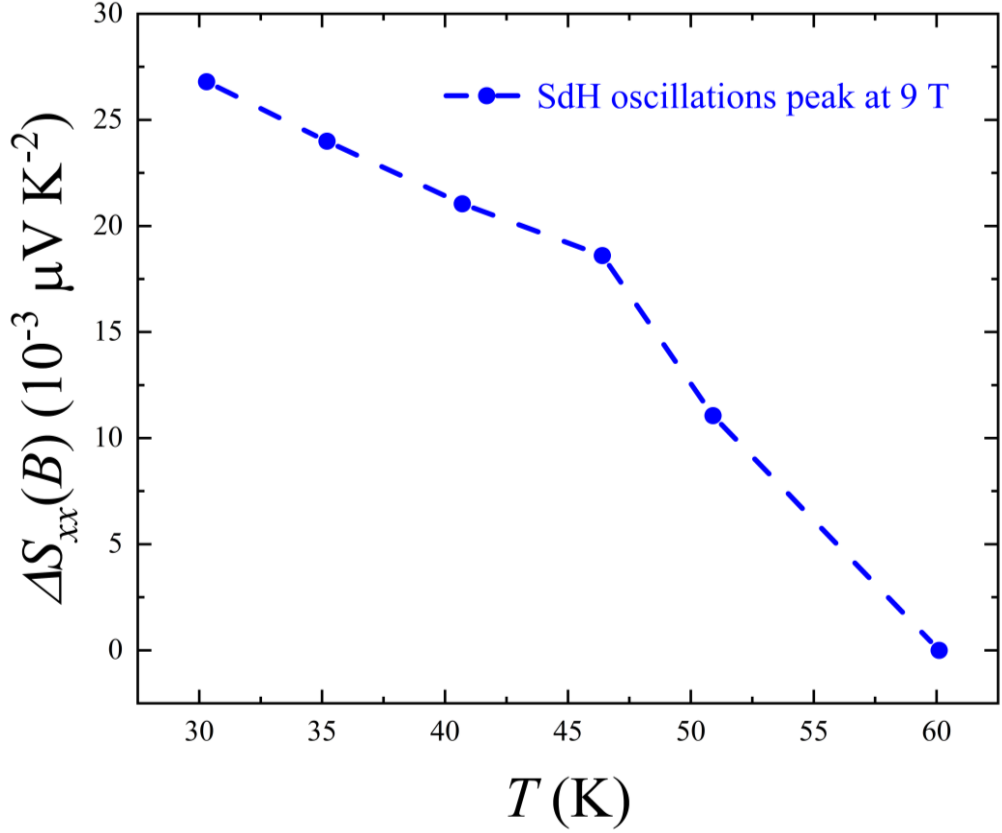

**Figure S2** . (Color online) Temperature dependence of the peak height of quantum oscillations of the thermopower (ΔS/T) at 9 T. The peak height at 8.8 T with respect to the peak height at 5.2 T from the baseline was calculated using the Dingle term ($R_D$) from the Lifshitz-Kosevich formula. The $T_D$ for the higher temperatures was estimated in a similar way as for the 10 K temperature (see main text).

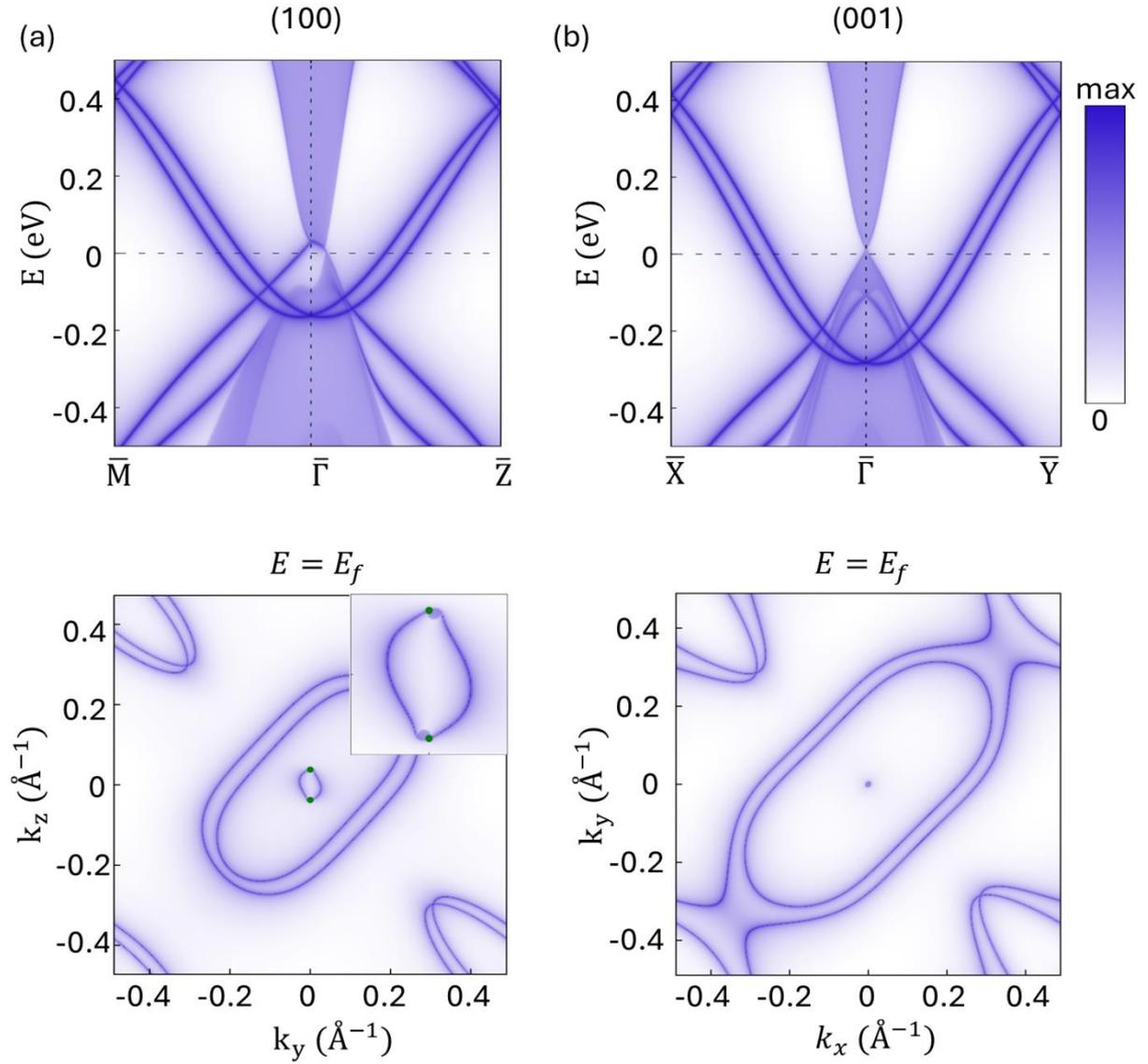

**Figure S3**. (Color online) The electronic band structure of the surface is projected on (a) (100) along $\bar{M}\to\bar{\Gamma}\to\bar{Z}$ and (b) (001) along $\bar{X}\to\bar{\Gamma}\to\bar{Y}$. The isoenergic Fermi-surface contour is displayed in the bottom panel, Dirac points are located with green circle.

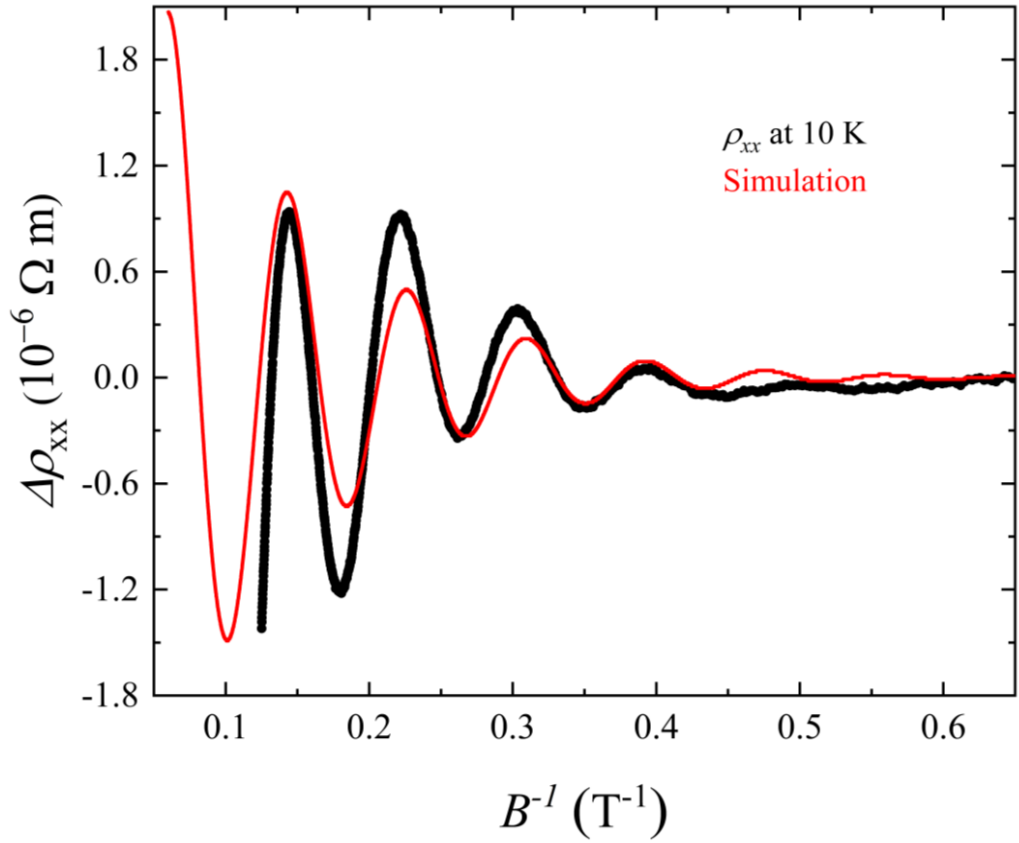

**Figure S4**. (Color online) (a) Quantum oscillations after subtracting the smooth 3$^{rd}$ order polynomial background in the field range of 1.5 to 8 T from $\rho_{xx}$(B) at 10 K temperature. Solid Red line presents the simulation of $\Delta\rho_{xx}$ vs $B^{-1}$ using the Lifshitz-Kosevich formula in the field range of 1.5 to 14.5 T. The parameters were used for the simulation are frequency $f_{SdH}$ ~ 12 T, effective mass $m^* \approx 0.01\ m_e$, Dingle temperature $T_D$ = 11 K, and additional phase shift $\beta \pm \delta = 0.7$. These parameters have been extracted from the fitting procedure described in the main text.